\documentclass[10pt]{emulateapj}
\usepackage{apjfonts}
\usepackage{psfig}
\usepackage{float}

\usepackage{graphics}
\usepackage{amssymb,amsmath}
\usepackage{color}

\newcommand{\chisq}{\ensuremath{\chi^2}}
\newcommand{\msun}{\ensuremath{\rm{M}_\odot}}
\newcommand{\rsun}{\ensuremath{\rm{R}_\odot}}

\newcommand{\kms}{\ensuremath{\rm{km \: s}^{-1}}}

\newcommand{\swift}{{\it Swift}}

\newcommand{\sneia}{SNe Ia}
\newcommand{\snia}{SN Ia}

\newcommand{\powerfittoall}{\ensuremath{ 2456062.30^{+ 0.24 }_{- 0.26 } }} 
\newcommand{\powerfittrise}{\ensuremath{ 19.50^{+ 0.56 }_{- 0.56 } }} 
\newcommand{\discoveryafter}{\ensuremath{ 2.42^{+ 0.24 }_{- 0.26 } }}

\newcommand{\WThreeeSigma}{\ensuremath{ 0.63} \AA}

\newcommand{\MassLimitPlotted}{\ensuremath{ 7.8 \times 10^{-3}} \msun}
\newcommand{\MassLimit}{\ensuremath{ 7.8 \times 10^{-3}} \msun}

\newcommand{\Binning}{\ensuremath{ 2.0} \AA}
\newcommand{\SmoothWidth}{\ensuremath{          250} \AA}

\newcommand{\FluxLimitPlotted}{\ensuremath{ 3.28 \times 10^{-17}} erg s\ensuremath{^{-1}} cm\ensuremath{^{-2}}}

\newcommand{\toconservup}{\ensuremath{-1.7}} 
\newcommand{\toconservdw}{\ensuremath{-2.6}} 
\newcommand{\rmax}{\ensuremath{0.16}} 
 
\newcommand{\rmaxdark}{\ensuremath{0.24}}

\begin{document}

\title{Strong Evidence Against A Non-Degenerate Companion in SN~2012cg}  
\shorttitle{No Companion Seen in SN~2012cg} 
\shortauthors{Shappee, Piro, Stanek, Patel, Margutti, Lipunov, \& Pogge}

\author{
{B.~J.~Shappee}\altaffilmark{1,2,3}, 
{A.~L.~Piro}\altaffilmark{2},
{K.~Z.~Stanek}\altaffilmark{4,5},
{S.~G.~Patel}\altaffilmark{2},
{R.~A.~Margutti}\altaffilmark{6,7},
{V.~M.~Lipunov}\altaffilmark{8,9},
and 
{R.~W.~Pogge}\altaffilmark{4,5}
}

\email{shappee@hawaii.edu}

\altaffiltext{1}{Institute for Astronomy, University of Hawai'i, 2680 Woodlawn Drive, Honolulu, HI 96822,USA}
\altaffiltext{2}{Carnegie Observatories, 813 Santa Barbara Street, Pasadena, CA 91101, USA}
\altaffiltext{3}{Hubble, Carnegie-Princeton Fellow}
\altaffiltext{4}{Department of Astronomy, The Ohio State University, 140 West 18th Avenue, Columbus, OH 43210, USA}
\altaffiltext{5}{Center for Cosmology and AstroParticle Physics (CCAPP), The Ohio State University, 191 W.\ Woodruff Ave., Columbus, OH 43210, USA}
\altaffiltext{6}{Center for Interdisciplinary Exploration and Research in Astrophysics (CIERA) and Department of Physics and Astronomy, Northwestern University, Evanston, IL 60208, USA}
\altaffiltext{7}{Center for Cosmology and Particle Physics, New York University, 4 Washington Place, New York, NY 10003, USA}
\altaffiltext{8}{M.V.Lomonosov Moscow State University, Physics Department, Leninskie gory, GSP-1, Moscow, 119991, Russia}
\altaffiltext{9}{M.V.Lomonosov Moscow State University, Sternberg Astronomical Institute, Universitetsky pr., 13, Moscow, 119234, Russia}

\date{\today}

\begin{abstract}

Even though SN~2012cg is one of the best-studied Type Ia Supernovae to date, the nature of its progenitor system has been debated in numerous studies.  Specifically, it is difficult to reconcile recent claims of the detection of a $\sim 6$ \msun{} main-sequence companion with recent deep, late-time H$\alpha{}$ flux limits. In this study we add three new constraints:  1) We analyze new high-signal-to-noise, nebular-phase, LBT/MODS spectrum of SN~2012cg and place an upper limit on the amount of low-velocity, solar-abundance material removed from a possible companion of $ < \MassLimit{}$.  2) We use \swift{} X-ray observations to constrain the preexplosion mass-loss rate to be $\dot M<10^{-6}\,\rm{M_{\sun}yr^{-1}}$ for $v_\textrm{w}=100\,\rm{km\,s^{-1}}$.  3) We carefully reanalyze a prediscovery MASTER image and, with published light curves of SN~2012cg, we estimate the time of first light and conservatively constrain the radius of a Roche-lobe overflowing companion to be $< \rmaxdark{}$ \rsun{}.  These observations disagree with a large nearby companion, and, when considered with other studies of SN~2012cg's progenitor system, essentially rule out a non-degenerate companion.

\end{abstract}
\keywords{supernovae: specific (SN\,2012cg)}

\section{Introduction}
\label{sec:introduc}

The physical nature of the progenitor systems of Type Ia Supernovae (\sneia{}) remains largely elusive despite decades of work (for a review see \citealp{wang12}). It is generally accepted that \sneia{} result from the thermonuclear explosion of a carbon-oxygen white dwarf (WD; \citealp{hoyle60}) in a close binary system, but there are still two competing classes of models. In the double-degenerate (DD) model, the WD's companion is another WD where the binary merges either due to the removal of energy and angular momentum from the system by gravitational radiation \citep{tutukov79, iben84, webbink84} or due to the perturbations of a third (e.g., \citealp{thompson11, katz12, shappee13c, antognini14}) or fourth \citep{pejcha13} body.  In the single-degenerate (SD) model, the WD's companion is a non-degenerate star \citep{whelan73, nomoto82}.  In most current models, the WD accretes matter from the non-degenerate companion until the WD undergoes unstable runaway nuclear burning. However, many current explosion simulations of both SD (e.g. \citealp{kasen09}) and DD (e.g. \citealp{pakmor12}) progenitors can match the observable signatures of SNe Ia around $B$-band maximum light ($t_{B {\rm max}}$), and observational searches for distinguishing characteristics between these two models have proven difficult.  

Despite these challenges, significant progress has been made during the past decade, and some classes of SD models have been observationally ruled out as the dominant channel (e.g., \citealp{bianco11, lipunov11, chomiuk16, maguire16}).  However, there are still candidate SD progenitor systems (e.g. U Sco and V445 Pup; \citealp{li11}).   Additionally, in the past few years, detailed studies of individual nearby \sneia{} have proven particularly fruitful.  A prime example is SN~2011fe \citep{nugent11ATel}, the brightest SN Ia in almost 40 years at just 6.4 Mpc away \citep{shappee11}, which was discovered by the Palomar Transient Factory (PTF; \citealp{law09}) less than one day after explosion.  For SN~2011fe, none of the observational signatures expected for SD channels have been seen \citep{horesh12, chomiuk12, margutti12, brown12, shappee13}, while early-light-curve properties \citep{piro10, nugent11, bloom12} and direct evidence for nuclear synthetic yields argue for a DD channel \citep{shappee17}.  However, there might be multiple channels for producing normal SNe Ia (e.g., \citealp{maguire13, yamaguchi15}), so additional well-studied \sneia{} are needed. 

In the SD model, the companion is expected to be struck by the SN ejecta soon after explosion.  This feature of the SD model has proven to be a fruitful observational test.  First, the interaction of the SN ejecta with the companion will modify the rising light curve at early times. Such a signature is dependent on the viewing angle, and the strongest effect will occur when the companion lies along the line of sight between the observer and the SN. At a fixed viewing angle, this emission scales proportionally with the companion's radius ($R_\textrm{c}$), which allows early-time observations to constrain the properties of the companion \citep{kasen10}.  Second, material from the companion will be stripped when struck by the SN ejecta (e.g., \citealp{wheeler75, marietta00, meng07, pakmor08, pan12, liu12}).  \citet{pan12} and \citet{liu12} use hydrodynamic simulations to show that $\sim 0.1 \-- 0.2$ \msun{} of solar-metallicity material is expected to be removed from main-sequence (MS) companions and that this material will be embedded in low-velocity supernova debris with a characteristic velocity of $\lesssim$1000 \kms{}.  Initially, this material will be hidden by the higher-velocity, optically thick, iron-rich ejecta, but will then appear in late-time, nebular-phase spectra ($\gtrsim 250$ days; \citealp{mattila05}) as the higher-velocity ejecta become optically thin.  Finally, this impact is expected to affect the companion's future properties \citep{podsiadlowski03, pan12b, shappee13b}.  

The Lick Observatory Supernova Search (LOSS; \citealp{filippenko01}) discovered the nearby SN~2012cg in NGC~4424 ($z = 0.001458 +\pm 0.000007$; \citealp{kent08}) on 2012 May 17.2 UT (MJD = 56064.22) and quickly announced it to the community \citep{cenko12ATEL}.  \citet{munari12} determined $t_{B {\rm max}}$ was 2012 June 4.5 UT (MJD=56082.0).  At just $15.2 \pm 1.9$ Mpc ($\mu = 30.90 \pm 0.3$ mag; \citealp{cortes08}), SN~2012cg has been an ideal candidate for extensive follow-up with multiwavelength studies in the radio \citep{chomiuk12ATEL, chomiuk16}, the far-infrared (far-IR) with Herschel \citep{johansson13}, the IR with the Hubble Space Telescope (HST; \citealp{amanullah15}), the optical from the ground \citep{silverman12, munari12, yamanaka14, amanullah15, maguire16, marion16} and HST \citep{amanullah15, graur16}, the ultraviolet (UV) with \swift{} \citep{marion16} and HST \citep{amanullah15}, and X-rays with \swift{} \citep{margutti12ATel}.  

However, even though SN~2012cg is one of the best-studied \sneia{} to date, the nature of its progenitor remains debated in a number of studies:
\\* (1) \citet{silverman12} claim that their early-time observations from the 0.76m Katzman Automatic Imaging Telescope (KAIT) rule out a $\sim 1$ \msun{} ($a \approx 2$ \rsun{}) main-sequence companion for certain viewing angles and red-giant companions ($a \approx 400$ \rsun{}) for almost all viewing angles.  
\\* (2) \citet{marion16} use the same KAIT observations with additional observations from the F.~L.~Whipple Observatory 1.2m (FLWO), the Las Cumbres Observatory Global Telescope Network (LCOGT; \citealp{brown13}), \swift{}, and ROTSE-IIIb \citep{akerlof03} to argue that they see a blue, early-time excess in the optical light curve consistent with the \citet{kasen10} models of the SN ejecta interacting with a $\sim 6$ \msun{} main-sequence companion.
\\* (3) \citet{chomiuk12ATEL, chomiuk16} observed SN~2012cg with the Karl G.\ Jansky Very Large Array to place some of the deepest radio limits of any \sneia{}.\footnote{Only SN~2011fe and SN~2014J have deeper limits in the radio.}  The resulting limits on any preexisting wind are inconsistent with those observed for symbiotic binaries and most isolated red giants \citep{chomiuk16}.  Main-sequence B stars with $\gtrsim$6 M$_{\odot}$ may, however, be consistent with the radio limits so long that the mass transfer from the companion to the WD is very conservative.  
\\* (4) \citet{maguire13} obtained Very Large Telescope (VLT) XShooter spectra of SN~2012cg $-0.8$ and $+27.3$ days after $t_{B {\rm max}}$.  They detected blueshifted Na${\textrm{I}}$ D absorption that was taken to be tentative evidence for previous mass loss from an SD progenitor.  However, \citet{maguire13} note that recent DD models (e.g., \citealp{shen13, raskin13}) can also reproduce their observations.
\\* (5) \citet{maguire16} use VLT spectra from XShooter and FORS2 acquired $\sim 340$ days after $t_{B {\rm max}}$ to constrain late-time H$\alpha$ flux from SN~2012cg.  Using scaled models from \citet{mattila05}, \citet{maguire16} placed a strong limit of $< 0.005$  \msun{} on the amount of solar-metallicity material present at low velocity, $\lesssim$1000 \kms{}, in SN~2012cg.  This limit strongly disfavors a hydrogen-rich SD companion  in the progenitor system of SN~2012cg \citep{maguire16}. Additionally, \citet{maguire16} find no evidence for low velocity, $\lesssim$1000 \kms{}, He$_{\textrm{I}}$ emission at 5876, 10830, and 20590 \AA{} in their XShooter visible and near-IR (NIR) spectra. However, no theoretical model predictions of the peak luminosity for these He features exist and it is not clear if these observations rule out material removed from a He star companion.   
\\* (6) \citet{graur12ATEL} used archival HST preexplosion $WFPC2$ images of NGC~4424 to obtain magnitude limits on the progenitor system of $F606W > 25.5$ and $F814W > 25.8$ mag.  These limits rule out most supergiants as a possible binary companion.
\\* (7) \citet{graur16} observed SN~2012cg to very late times with HST and claim their observations are better described by the nuclear synthetic yields expected from a progenitor white dwarf with a mass near the Chandrasekhar limit. However, these observations are in a single long-pass filter making the measured decay sensitive to changes in the underlying spectral energy distribution.  \citet{graur16} also note that their observations are consistent with a light echo. 
\\* (8) \citet{liu16} attempted to reconcile these observational studies with theoretical predictions from binary evolution and population-synthesis calculations for a range of progenitor scenarios.  They find that either a SD progenitor with a 6 M$_{\odot}$ MS companion or a DD model with carbon-oxygen WD donor might be the progenitor of SN~2012cg, but both scenarios are in conflict with at least some of the reported observations. The most difficult observations to reconcile are the early-time light-curve detection from \citet{marion16} and the late-time hydrogen limits from \citet{maguire16}.

In this study, we add a new late-time H$\alpha$ constraint, a new X-ray constraint, and reanalyze a timely prediscovery photometric observation of SN~2012cg.  Our findings agree with the \citet{maguire16} H$\alpha$ limits and disagree with the \citet{marion16} claim of an excess in the early-time light curve from SN~2012cg.  When our constraints are considered with other studies of SN~2012cg's progenitor system, the available observations strongly favor a DD progenitor.  In Section \ref{sec:hydrogen}, we present our new deep nebular phase spectra and place a limit on the amount of hydrogen-rich, low-velocity material possible in SN~2012cg.  In Section \ref{sec:xray}, we describe and discuss \swift{} X-ray observations and limits.  In Section \ref{sec:lc}, we present the MASTER photometry, refit the early-time light curve, and place constraints on the presence of a non-degenerate companion.  Finally, in Section \ref{sec:sumamry}, we discuss the observational constraints on SN~2012cg and conclude that a DD model is the most likely progenitor of SN~2012cg. 

Throughout this paper we assume total reddening of SN~2012cg of $E(B-V) = 0.15$ and a ratio of total to selective absorption $R_{V}= 2.6$ \citep{amanullah15}.  \footnote{For completeness we note that \citet{marion16} assume $R_{V}= 3.1$.  This difference, however, is small and would not qualitatively affect the claims in either paper.}

\section{The Search for Companion Material}
\label{sec:hydrogen}

In this section, we place a deep and constraining limit on the presence of hydrogen-rich, low-velocity material in SN~2012cg. We obtained a high-signal-to-noise spectrum of SN~2012cg 286 days after $t_{B {\rm max}}$ (MJD = 56368.3) using the Large Binocular Telescope (LBT) + Multi-Object Double Spectrographs (MODS; \citealp{pogge10}). We obtained $8 \times 1200$s exposures through a $1\arcsec$ wide slit.  The spectra reduction was the same as described in \citet{shappee13}. We estimate that our absolute flux calibration around H$\alpha$ to be accurate to $10 \%$ or better.

\begin{figure}[htp]
	\includegraphics[width=8.8cm]{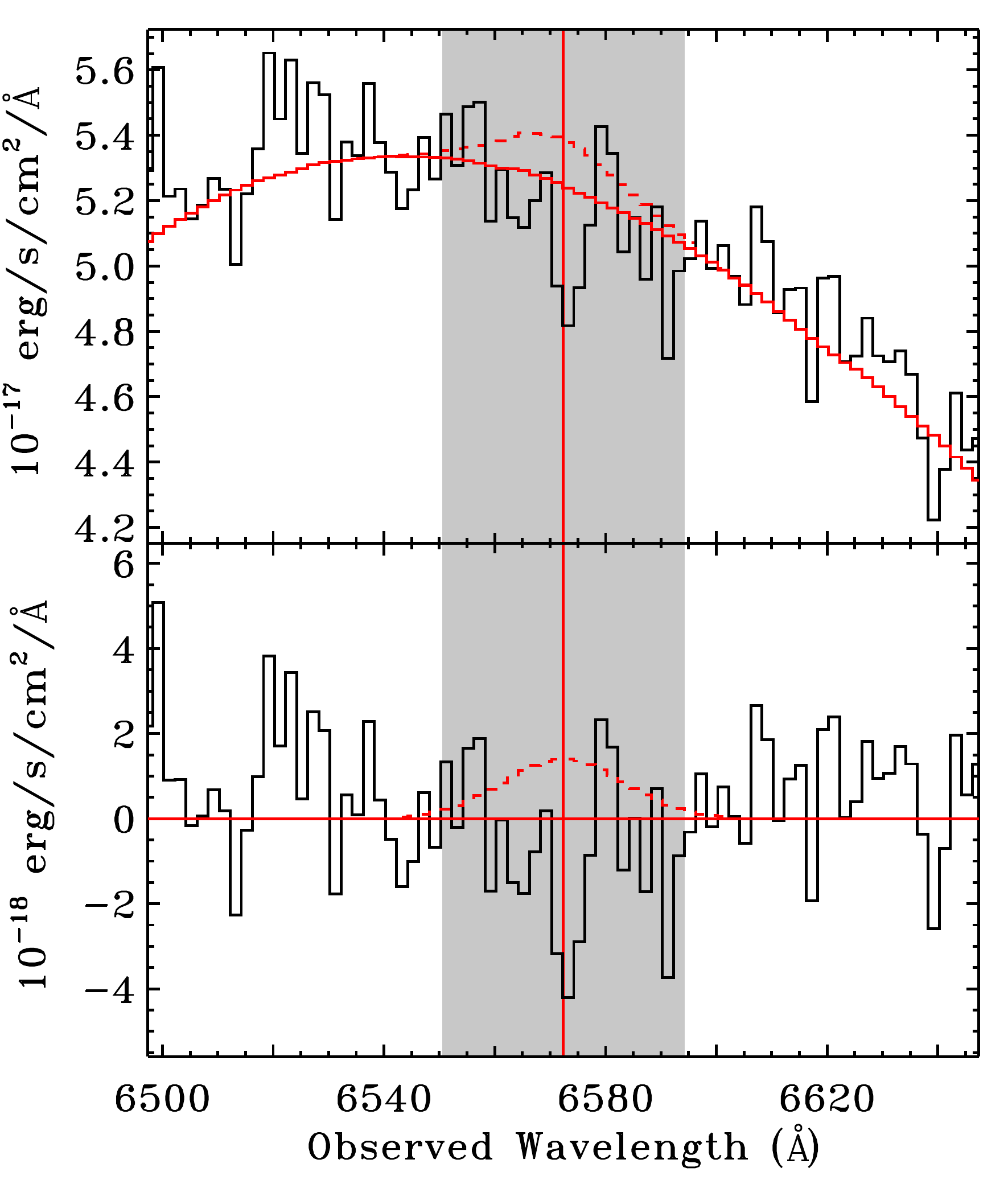}

	\caption{Nebular phase spectrum of SN~2012cg, illustrating our H$\alpha$ flux limit of \FluxLimitPlotted{}. The rest wavelength of H$\alpha$ is indicated by the vertical red line, and the gray shaded region shows where hydrogen emission would be expected ( $\pm 1000$ \kms{} $= \pm 22 $ \AA{} about H$\alpha$).  Adopting the models of \citet{mattila05}, these limits translate into a $\lesssim$\MassLimitPlotted{} limit on the amount of solar-abundance material stripped from the companion.  Note that a weak, narrow H$\alpha$ absorption line, likely from the host galaxy, might be present in the spectrum.  {\it Top panel:} The SN spectrum binned to the approximate spectral resolution (\Binning{}; black solid); smoothed continuum (solid red); and smoothed continuum with H$\alpha$ limit added (dashed red) are shown.  {\it Bottom panel:} SN spectrum with smoothed continuum subtracted (solid black) as compared to the H$\alpha$ limit (dashed red) are shown.  A horizontal solid red line at zero is shown for reference.}
	\label{fig:halpha}
\end{figure}

To place an H$\alpha$ limit from the spectrum of SN 2012cg we followed the methods presented in \citet{leonard07} as described in \citet{shappee13}.  Briefly, we first smooth the spectrum on a scale much larger than the expected width of an H$\alpha$ feature.  Varying the smoothing width from 100 - \SmoothWidth{} lead to no significant differences. We then subtract off this smoothed spectrum and look for any excess flux in the residuals around H$\alpha$. Our nebular phase spectrum and continuum fit in the vicinity of H$\alpha$ are shown in the top panel of Figure \ref{fig:halpha} binned to their approximate spectral resolution. We searched for H$\alpha$ emission within $\pm 1000$ \kms{} ($\pm 22 $ \AA) about H$\alpha$ at the redshift of NGC~4424. We note that this $1000$ \kms{} width is wider and thus more conservative than the \citet{mattila05} analysis. There is no evidence for any H$\alpha$ emission in the spectrum.

To quantify our limit we use equation 1 of \citet{leonard07} leading to a $3\sigma$ upper limit on the equivalent width of $W_\lambda(3\sigma) = $ \WThreeeSigma{} assuming a full-width at half-maximum of the H$\alpha$ spectral feature is $W_{\textrm{line}} = 22$ \AA{}.

\begin{figure*}[htp]
	\includegraphics[width=18.0cm]{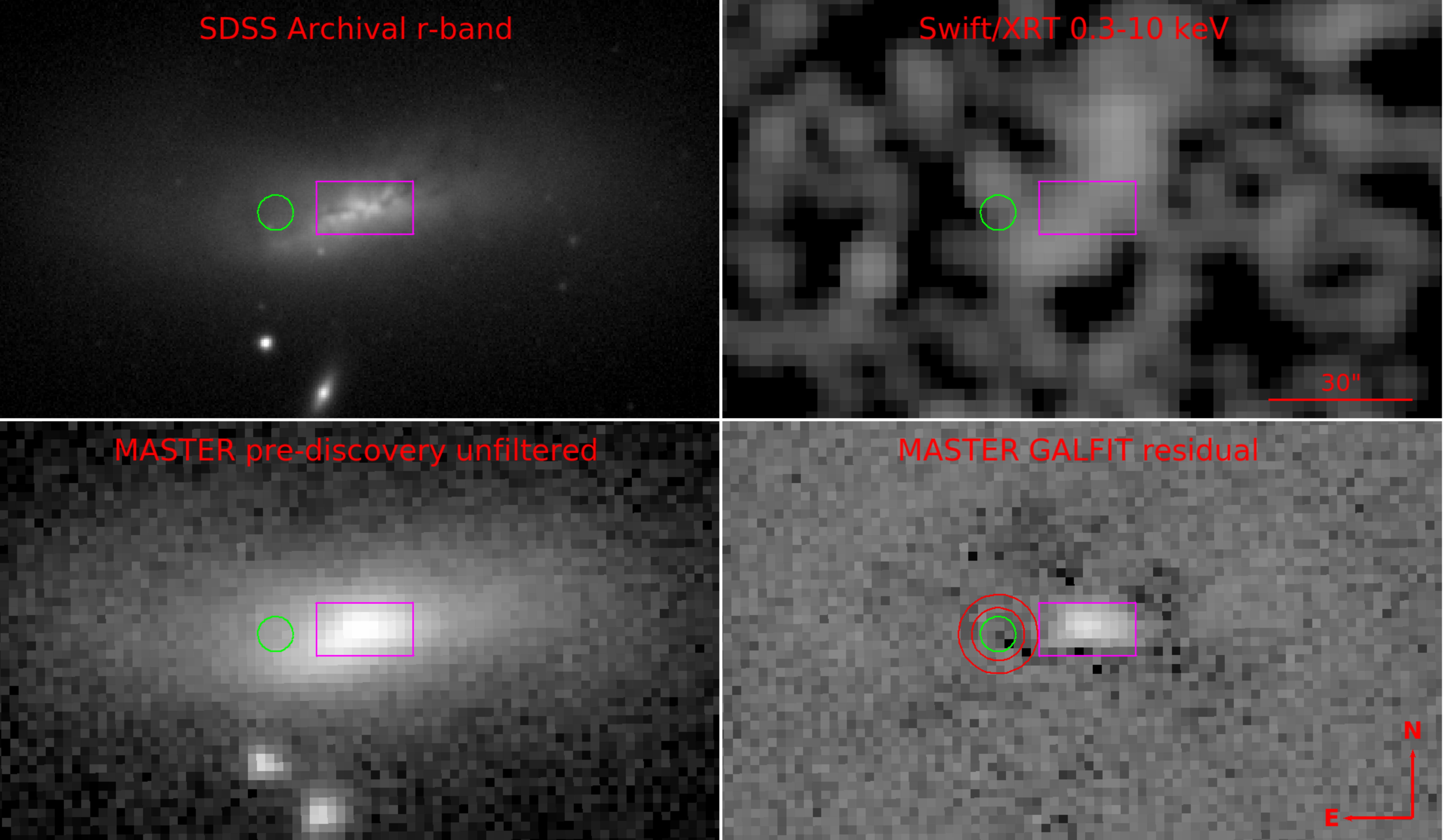}

	\caption{Optical and X-ray images at the location of SN~2012cg.  {\it Top left panel:} SDSS $r$-band preexplosion image.  {\it Top right panel:} \swift{}/XRT 0.3-10 keV X-ray image.  {\it Bottom left panel:} MASTER prediscovery unfiltered image. {\it Bottom right panel:} GALFIT residual image of MASTER prediscovery unfiltered image.  The green circles and the red annulus indicate the aperture and sky region, respectively, used for the MASTER's photometry, centered on the SN position. The purple rectangle shows the masked central region used in the GALFIT fit.}
	\label{fig:image}
\end{figure*}


To translate this equivalent width limit into a constraint on the amount of material stripped from a non-degenerate companion, we follow \citet{leonard07} and \citet{shappee13} by scalling from the results of \citet{mattila05}. After accounting for the distance to NGC~4424 and the measured extinction of SN~2012cg, 
we place an upper limit on the amount of solar-abundance material in SN~2012cg of \MassLimit{}. This limit agrees with the $< 0.005$  \msun{} limit reported by \citet{maguire16}, who also scale from the \citet{mattila05} 380-day model. However, because our  observation was acquired earlier (286 vs. $\sim 340$ days after $t_{B {\rm max}}$), when a given mass of material is expected to be more luminous, we view our limit as even more conservative.

\section{Swift X-ray Limits}
\label{sec:xray}

In this section, we analyze \swift{} X-ray data obtained for SN~2012cg and place limits on the mass-loss rate and wind velocity from its progenitor system. X-ray emission in a young SN originates from the SN shock interaction with the circumstellar medium (CSM). It directly depends on both the properties of the SN explosion (i.e. shock velocity, ejecta mass) \emph{and} the properties of the environment (i.e. the amount of mass previously ejected by the stellar progenitor). Depending on these properties, different emission processes are expected to contribute to the observed X-rays, in particular: (i) Inverse Compton (IC) scattering of photospheric optical photons off of relativistic electrons accelerated by the SN shock, (ii) synchrotron emission from the same population of accelerated electrons, and (iii) thermal bremsstrahlung.  In the low-density regime expected around Type Ia SN progenitors ($\dot{M}\lesssim 10^{-9}-10^{-4}$ M$_{\odot}$~yr$^{-1}$; \citealp{chomiuk16}), IC emission has been shown to dominate the X-ray emission at $t\lesssim40$ days (\citealp{margutti12}; see \citealp{chevalier06} for a review).  In Section \ref{sec:XRTObs} we describe the \swift{} observations, and we constrain the 0.3-10 keV X-ray flux.  In Section \ref{sec:massloss}, we place limits on SN~2012cg's progenitor system mass-loss rate.

\subsection{\swift-XRT Observations}
\label{sec:XRTObs}

The X-Ray Telescope (XRT, \citealt{burrows05}) on board  the \swift{} satellite \citep{gehrels04} started observing SN~2012cg on 2012 May 18 UTC ($MJD = 56065.8$) and carried out an extensive follow-up campaign.  Here we concentrate on observations obtained within $\sim 30$ days of explosion ($MJD \leq 56088.2$), which offer the best opportunity to constrain the presence of X-ray radiation originating from IC emission. The XRT data were analyzed using HEASOFT (v6.15) and corresponding calibration files. We used standard filtering and screening criteria.   

Combining the 24.7 ks of \swift/XRT data collected in this time interval, we find no evidence for X-ray emission from a point source at the position of SN~2012cg, with a 3$\sigma$ count-rate limit of $1.1\times 10^{-3}\,\rm{c/s}$ (0.3-10 keV).  Unfortunately, unresolved diffuse X-ray emission from the host galaxy prevents us from obtaining a deeper limit.  Figure~\ref{fig:image} shows the \swift{}/XRT 0.3-10 keV X-ray image.  The Galactic neutral hydrogen column density in the direction of SN\,2012cg is $N(H)_{MW}=1.5\times 10^{20}\,\rm{cm^{-2}}$ \citep{kalberla05}. From the measured total reddening and assuming a Galactic dust-to-gas ratio, we infer an intrinsic $N(H)_{int}\sim5.5\times 10^{20}\,\rm{cm^{-2}}$. With these parameters and adopting an absorbed power-law spectral model with photon index $\Gamma=2$ (see Section \ref{sec:massloss}), the count-rate limit above translates into an unabsorbed flux limit $F<5.1\times 10^{-14}\,\rm{erg\,s^{-1}cm^{-2}}$, corresponding to $L<1.4\times 10^{39}\,\rm{erg\,s^{-1}}$ (0.3-10 keV).

\subsection{Constraints on the progenitor system mass-loss rate from X-ray observations}
\label{sec:massloss}

To place constraints on the progenitor-system mass-loss rate, we adopted the generalized IC formalism developed in \citet{margutti12}. We assumed the following: 1) The SN outer density structure is $\rho_{SN}\propto R^{-n}$ with $n\sim10$ \citep{matzner99}. 2) The electrons are accelerated by the SN shock in a power-law distribution $n(\gamma)\propto \gamma^{-p}$ with $p=3$, as found from radio observations of hydrogen-stripped SN explosions (e.g. \citealt{soderberg06}).  Thus we expect a photon index $\Gamma=2$. 3) The fraction of post-shock energy density in relativistic electrons is $\epsilon_e=0.1$ where the limit on the density scales as $(\epsilon_e/0.1)^{-2}$. (See Appendix~A in \citealt{margutti12}.) 

We next estimated the bolometric luminosity of SN~2012cg, which is required to translate the X-ray flux limit into a physical limit on mass loss in the progenitor system.  We used the UV through NIR photometry published by \citet{marion16}. For each epoch, we corrected for the observed extinction and distance to NGC~4424 and then integrated the flux density over the wavelength range of the observations. We lack NIR measurements after 9 days past $t_{B {\rm max}}$, so we linearly extrapolated the NIR photometry and added a conservative error estimate. For a 10,000 K blackbody, the amount of flux falling outside the wavelength range covered by the photometry is $\sim7$\%.  We adopt this correction to reconstruct the bolometric luminosity of SN~2012cg around the time of the optical peak.

By combining the bolometric light curve with the X-ray limits calculated in Section~\ref{sec:XRTObs}, we can constrain the environment density around SN~2012cg. For an ISM medium where $\rho_{\textrm{CSM}}=const$, we derive $\rho_{\textrm{CSM}}<10^5\,\rm{cm^{-3}}$. However, a star that has been losing material at constant rate $\dot M$ gives rise to a wind-like CSM where $\rho_{\textrm{CSM}}=\dot M/(4\pi r^2\,v_{\textrm{w}})$, where $v_\textrm{w}$ is the wind velocity. For a wind-like medium, we infer $\dot M<10^{-6}\,\rm{M_{\sun}yr^{-1}}$ for $v_\textrm{w}=100\,\rm{km\,s^{-1}}$.  

\begin{figure}
	\includegraphics[width=8.8cm]{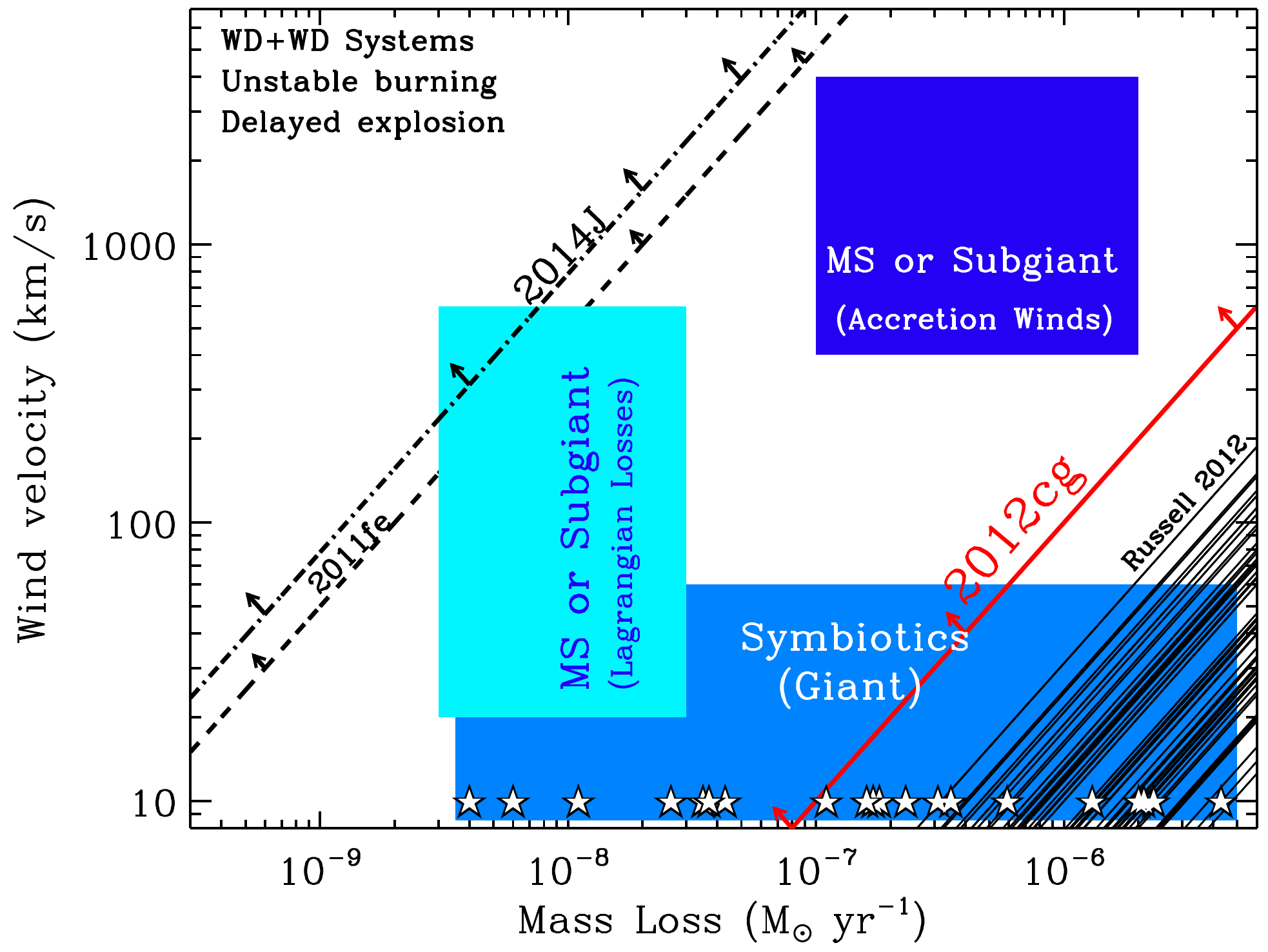}

	\caption{Limits on mass loss rates as a function of wind velocity.  The limit for SN~2012cg from the \swift{}/XRT flux limits are shown by the solid red line.  For comparison, limits on SN~2012cg's progenitor system mass-loss rate from \swift{}/XRT X-ray non-detections are shown.  The limits from SN~2011fe \citep{margutti12} and SN~2014J \citep{margutti14} are  represented by the dashed and dot-dashed lines, respectively.  Other \swift{} limits are indicated by lines in the lower right of the figure \citep{russell12}.  Regions for possible \sneia{} progenitor systems are also indicated. White stars indicate the measured mass loss rates of Galactic symbiotic systems for an assumed $v_\textrm{w} = 10$ km s$^{-1}$ }
	\label{fig:SD}
\end{figure}

In Figure~\ref{fig:SD}, we compare these limits to published limits \citep{margutti12, margutti14, russell12} and candidate \sneia{} progenitor systems.  We find that, because of the X-ray emission from the host galaxy, our \swift{}/XRT limits are not very constraining.  We can rule out only a fraction of symbiotic progenitor systems, and we would not expect to detect signatures from MS or subgiant companions.

\section{Early-time Light Curve and Companion Constraints}
\label{sec:lc}

If the progenitor of an \snia{} is a WD accreting from a non-degenerate companion, then the SN ejecta will interact with the companion and potentially produce an observable feature in the rising light curve at early times. Such a signature is dependent on the viewing angle, with the strongest effect occurring when the companion lies along the line of sight between the observer and the SN. At a fixed viewing angle, this emission scales proportionally with $R_\textrm{c}$.  

In this section, we constrain the radius of a possible companion in the progenitor system of SN~2012cg using its very early-time light curve.  In Section~\ref{sec:marionlcs}, we introduce light curves from \citet{marion16}. In Section~\ref{sec:preobs} we present three prediscovery observations that were not included in the analysis of \citet{marion16}.  In Section~\ref{sec:lcfit}, we closely follow the methods of \citet{shappee16a} to fit the early-time light curves and constrain the time of first light ($t_{\textrm{first}}$).  Finally, in Section ~\ref{sec:progenitor}, we use these observations, our estimate of $t_{\textrm{first}}$, and the interaction models of \citet{kasen10} to place stringent limits on $R_\textrm{c}$.

\subsection{KAIT, LCOGT, FLWO, \swift{}, and ROTSE Observations}
\label{sec:marionlcs}

We use the optical photometry published by \citet{marion16} from KAIT ($B$, $V$, $R$, and $I$), FLWO ($U$, $B$, $V$, $r$, and $i$), LCOGT ($B$ and $V$), ROTSE-IIIb (unfiltered but calibrated to $r$-band), and the UV/Optical Telescope (UVOT; \citealp{roming05}) on board \swift{} ($u$, $b$, and $v$).  The NIR observations from the Peters Automated Infrared Imaging Telescope presented in \citet{marion16} were not acquired early enough to be useful for fitting $t_{\textrm{first}}$ or constraining a possible companion, and they are not used in this section.  Additionally, there were only 2, 2, and 3 \swift{} UV observations in the $uvw2$, $uvm2$, and $uvw1$ filters within 6 days of discovery, respectively\footnote{We note that only two of the first three $uvw1$ observations are shown in Figure~5 of \citet{marion16}.}, so these observations are also not useful for constraining $t_{\textrm{first}}$ but are discussed in Section~\ref{sec:sumamry}.   

We transformed the \swift{} $b$ and $v$ observations to Bessel $B$ and $V$ bands, respectively, using the UVOT-Johnson $UBV$ color corrections\footnote{\url{http://heasarc.gsfc.nasa.gov/docs/heasarc/caldb/swift/docs/uvot/uvot\\\_caldb\_coltrans\_02b.pdf}} derived from the Pickles stars.  These transformations are small, $< 0.03$ mag, and do not qualitatively affect our results.  We transformed the \swift{} $u$ into AB-magnitudes using the transformations presented on the \swift{} website\footnote{\url{http://swift.gsfc.nasa.gov/analysis/uvot\_digest/zeropts.html}} and, compare it directly Sloan $u$ data because the filter response functions are somewhat similar.  However,  when interpreting the $u$-band observations care should be taken for two reasons.  First, SNe Ia have strong Ca II H\&K lines (3945 \AA) that rapidly change in strength and velocity in their early-time spectra.  Furthermore, Ca II H\&K is located right on the edge of $u$-band.  Second, the Sloan $u$ and \swift{} $u$ are not the same filter and the earth's atmosphere substantially affects $u$.  Thus, a careful $S$-corrections are required to correct for the differences in the actual band passes which is beyond the scope of this paper.  Unfortunately, the ground-based $u$-band observations start too late to be useful on their own in constraining the presence of a potential companion.  Thus we include the \swift{} $u$ observations because they do not qualitatively affect our $t_{\textrm{first}}$ fits, were already shown in \citet{marion16}, and are potentially useful when comparing with the ground-based $u$-band photometry.  However, we strongly warn the reader about their reliability. 

\subsection{Prediscovery Observations}
\label{sec:preobs}

There have been three prediscovery observations reported in the literature:  
\\* (1) The previous KAIT observation of the host galaxy of SN~2012cg was on 2012 May 12.3 UTC ($MJD = 56059.3$), 4.9 days before discovery \citep{silverman12}.  
\\* (2) \citet{cortini12} report a 19.0 magnitude unfiltered upper limit using a Celestron C14 telescope + Starlight X-Press SXVR H9 camera on 2012 May 14.9 UTC ($MJD = 56061.9$), 2.6 days before discovery.  The Starlight X-Press SXVR H9 camera's efficiency peaks at 5400\AA\footnote{\url{https://www.sxccd.com/handbooks/SXVF-H9\%20handbook.pdf}}, and \citet{silverman12} report this limit as $R$ band.  Calibrating an unfiltered observation to a red filter is a conservative when placing constraints on a possible shock interaction because  the interaction is expected to be stronger at shorter wavelengths. 
\\* (3) As part of the ongoing MASTER survey \citep{lipunov10}, we acquired an unfiltered prediscovery image on 2012 May 15.79 UTC ($MJD = 56062.79$), 1.43 days before discovery.  \citet{lipunov12ATel} report a marginal detection of SN~2012cg at $R \approx 19$ mag in this image.  

In an attempt to verify the prediscovery MASTER detection, we reanalyze this prediscovery image.  Typically, we would use image subtraction to remove the host galaxy light with images taken at the same location and instrument without the SN present.  However, no additional images are available without SN~2012cg being present.  Instead, we used GALFIT \citep{peng02} to model the large-scale, 2-D light distribution of the galaxy.  A nearby star served as a PSF model for the Sersic fit.  Neighboring objects were also modeled and simultaneously fit, with initial guesses for the Sersic parameters obtained from SExtractor.  The complex structure of the host galaxy warranted two Sersic components, especially given the large residuals from a single-component fit.  Indeed, both components provided extended fits to the galaxy light distribution.  Additionally, we masked the central region so that the fit would be driven by the smooth extended light at large radii around the SN.   Figure~\ref{fig:image} presents the SDSS archival $r$-band image, the MASTER prediscovery image, and the GALFIT residual image.  

We then performed aperture photometry at the location of SN~2012cg using the IRAF {\it apphot} package and calibrated the results to $r$-band.  Again, calibrating an unfiltered magnitude to a red filter is a conservative choice.  There was no excess flux detected at the location of SN~2012cg, and we place a 3-sigma limit of $r > 19.7$  mag.  We note that the SDSS $r$-band image reveals a point source near the location of SN~2012cg. To verify that this source has not substantially affected our limit, we performed aperture photometry on the SDSS $r$-band image and found that this source has $r = 20.3 \pm 0.6$, which is significantly fainter than our upper limit.  Finally, to verify that any flux from the SN is not impacting the GALFIT fit, we injected artificial sources at the location of the SN in the reduced image, ran GALFIT on the resulting image, and then attempted to recover the source.  When injecting 18.5, 19.0 and 19.5 mag sources, we recover each source at $18.80 \pm 0.19$, $19.12 \pm 0.24$, and  $19.41 \pm 0.28$ mag, respectively.  These artificial source tests show that our method effectively removes the galaxy light while not significantly affecting the flux from point sources.

\subsection{Early-Time Light Curve Fit}
\label{sec:lcfit}

\begin{figure}
	\centerline{
		\includegraphics[height=13cm]{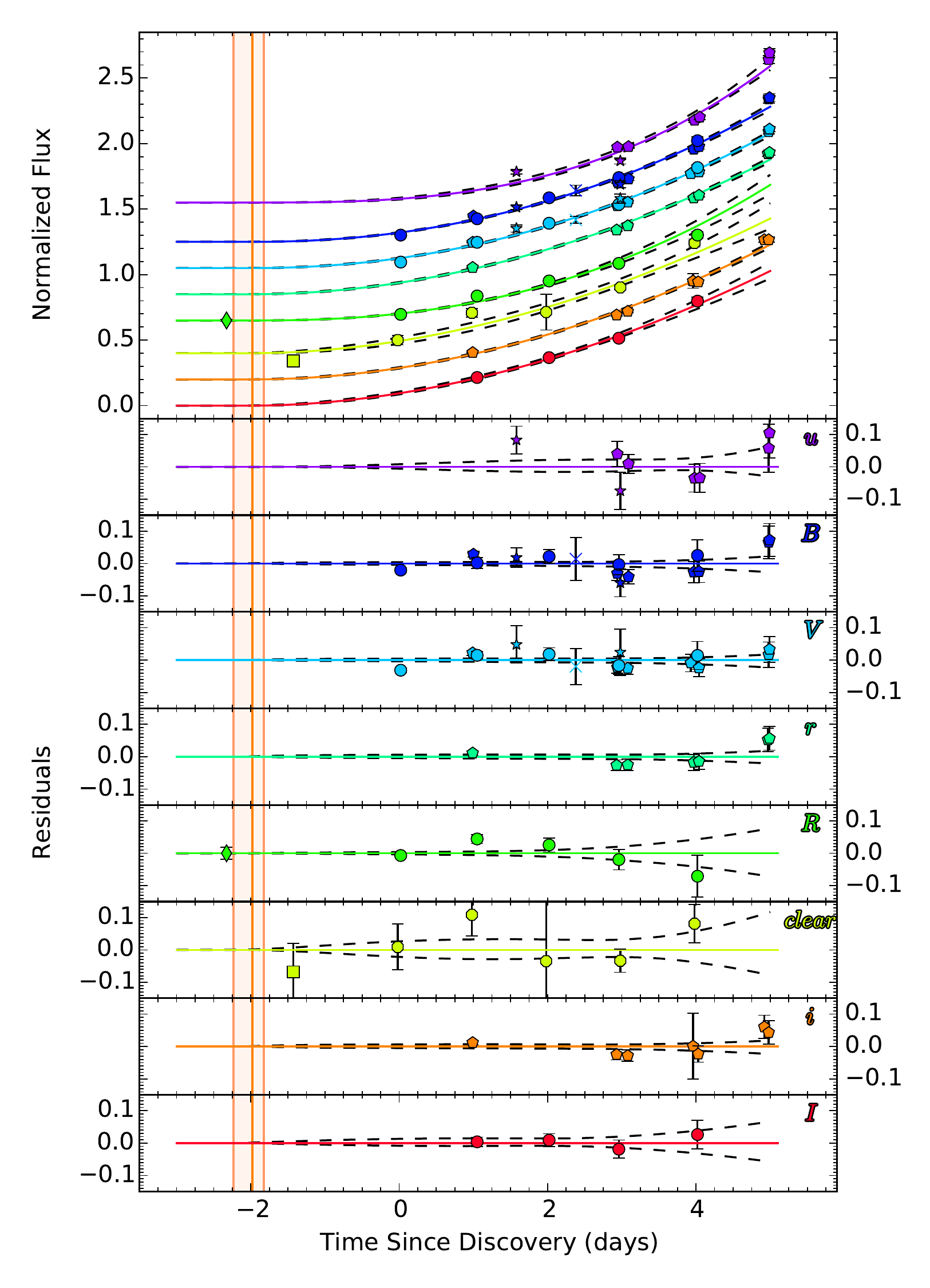}
	}
	\caption{Early-time light curves of SN~2012cg with power-law fits to observations obtained within $5$ days of discovery. Observations from KAIT (circles), FLWO (pentagons), LCOGT ('X's), \swift{} (stars), MASTER (square), ROTSE-IIIb (squares), and \citet{cortini12} (diamond) are shown with filled symbols. Error bars are shown but are sometimes smaller than the points. Colored solid lines are the best-fit power laws while dashed black lines show each fit's 1-sigma range.  The vertical orange line shows the best-fit $t_{\textrm{first}}$ and its uncertainty (see Section~\ref{sec:lcfit}).  {\bf Top panel:}  Light curves normalized by their flux at 5 days after discovery with an added constant.  {\bf Bottom panels:} Residuals from the best-fit power laws for each individual filter.}
	\label{fig:earlyfit}
\end{figure}

To determine $t_{\textrm{first}}$, we followed the methods presented in \citet{shappee16a}.  We modeled each band's photometry as an independent power law with index ($\alpha_{i}$) but forced all bands to have the same $t_{\textrm{first}}$.  This approach is motivated by \citep{piro16} who show in their Figure~2 that different band rise at different rates but begin to rise at the same time. We used the {\tt emcee} package \citep{foreman13}, a Python-based implementation of the affine-invariant ensemble sampler for Markov chain Monte Carlo (MCMC), to perform the fit to each light curve. As is done in \citet{goobar15} and \citet{shappee16a}, we fit the data in flux space (not logarithmic magnitudes) to allow the natural treatment of non-detections.  We fit all observations obtained within $5$ days of discovery, but the resulting fits were statistically unsatisfactory, at least partially because the photometry reported in \citet{marion16} had underestimated errors.  We first added 1\% error to all the photometry reduce the fits sensitivity to $S$-corrections required to translate between similar filters on different telescopes. This correction was not preformed in \citet{marion16} and is beyond the scope of this study.  We then scaled the errors for each filter so that \chisq{} per degree of freedom was 1 for the best fit for that filter. Each filter was scaled by a factor of 3 or less, except for $u$ and $R$ which were scaled by 3 and 5, respectively.  We then refit all the observations obtained within $5$ days of discovery.  The best-fit power laws for each filter and their corresponding 1-sigma uncertainties are shown in Figure~\ref{fig:earlyfit}.   We find that JD $t_{\textrm{first}} = \powerfittoall$, implying that SN~2012cg was discovered \discoveryafter{} days after $t_{\textrm{first}}$ and that $t_{\textrm{rise}} = t_{B {\rm max}} - t_{\textrm{first}} = \powerfittrise$ days.  The best-fit $\alpha$ parameters for each band are presented in Table~\ref{tab:lcfit}.  We note that the original fit (before scaling errors) is contained within our reported parameter ranges and the main result of increasing the errors is to broaden our uncertainty on $t_{\textrm{first}}$.

\begin{deluxetable}{cr}
\tablewidth{100pt}
\tabletypesize{\small}
\tablecaption{Fit Light Curve Parameters}
\tablehead{
\colhead{band} &
\colhead{$\alpha$}  }
 $u$ & $ 2.86^{+ 0.18 }_{- 0.24 } $ \\
 $B$ & $ 2.13^{+ 0.09 }_{- 0.15 } $ \\
 $V$ & $ 2.08^{+ 0.09 }_{- 0.13 } $ \\
 $r$ & $ 1.96^{+ 0.09 }_{- 0.12 } $ \\
 $R$ & $ 2.37^{+ 0.15 }_{- 0.20 } $ \\
 $clear$  &  $ 1.91^{+ 0.24 }_{- 0.31 } $ \\
 $i$ & $ 1.96^{+ 0.09 }_{- 0.11 } $ \\
 $I$ & $ 1.90^{+ 0.14 }_{- 0.15 } $ \\
 \enddata
\label{tab:lcfit}
\end{deluxetable}

\begin{figure}
	\centerline{
		\includegraphics[width=8.8cm]{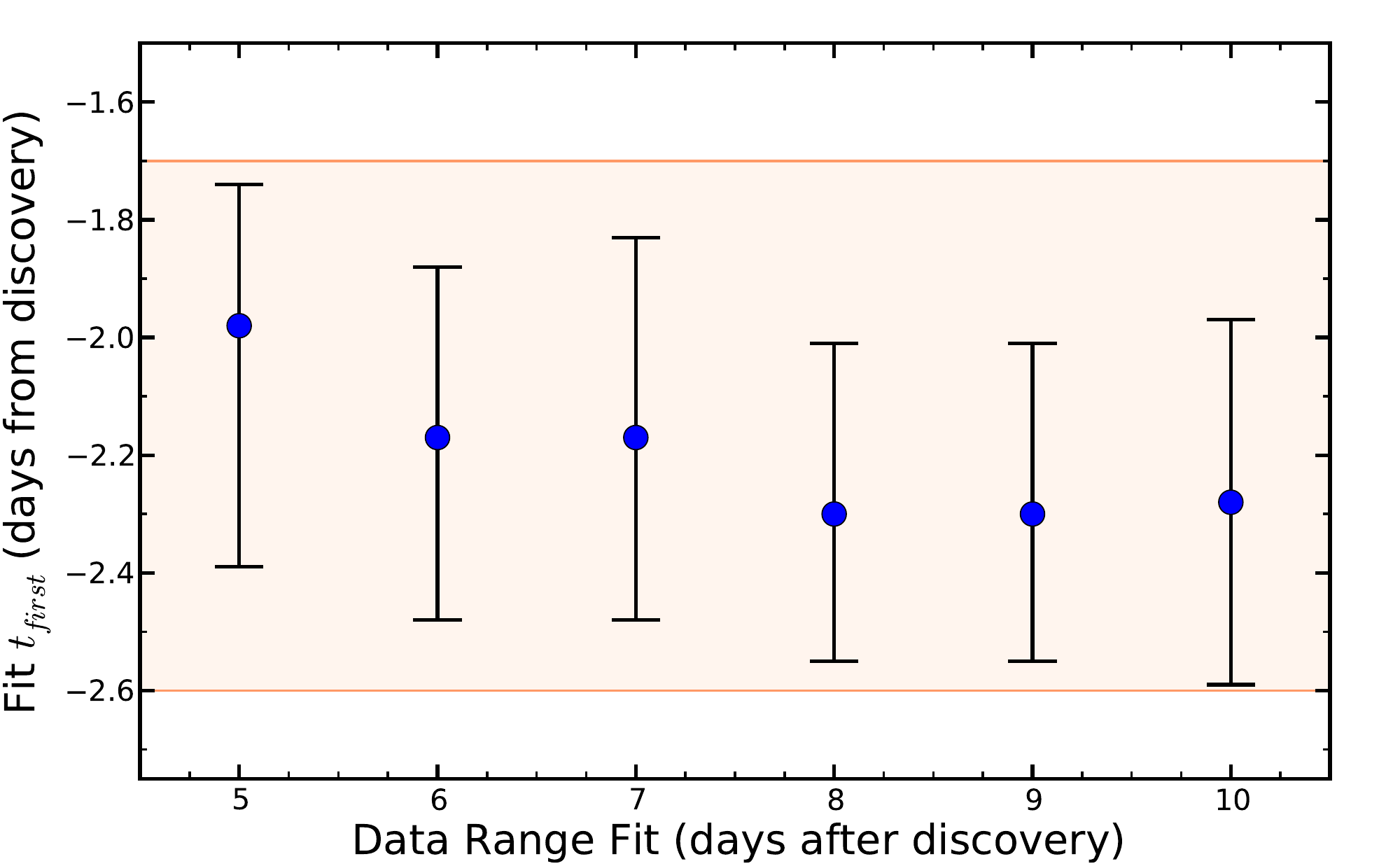}
	}
	\caption{Estimates of $t_{\textrm{first}}$ as a function of the data range fit and their 90\% confidence intervals.  The orange region shows the conservative estimate for $t_{\textrm{first}}$ we adopt.}
	\label{fig:tofit}
\end{figure}

However, the choice in the range of data to fit is somewhat arbitrary.  To estimate the systematic errors in our estimate of $t_{\textrm{first}}$, we varied the data range and repeated our MCMC fit to the light curves.  The resulting $t_{\textrm{first}}$ and 90\% confidence intervals when fitting  various data ranges are shown in Figure~\ref{fig:tofit}.  To be conservative, we adopt the range \toconservdw{} to \toconservup{} days from discovery for $t_{\textrm{first}}$ when placing progenitor constraints.

\subsection{Progenitor System Constraints}
\label{sec:progenitor}

To place constraints on the progenitor system of SN~2012cg, we first create absolute-magnitude light curves.  As elsewhere in the manuscript, we assume a total reddening of $E(B-V) = 0.15$ \citep{amanullah15}, a ratio of total-to-selective absorption $R_{V}= 2.6$, and a distance of $15.2 \pm 1.9$ Mpc ($\mu = 30.90 \pm 0.3$ mag).  We add the distance modulus errors in quadrature with the photometric errors.  

\begin{figure}[t]
	\centering
	\includegraphics[width=8cm]{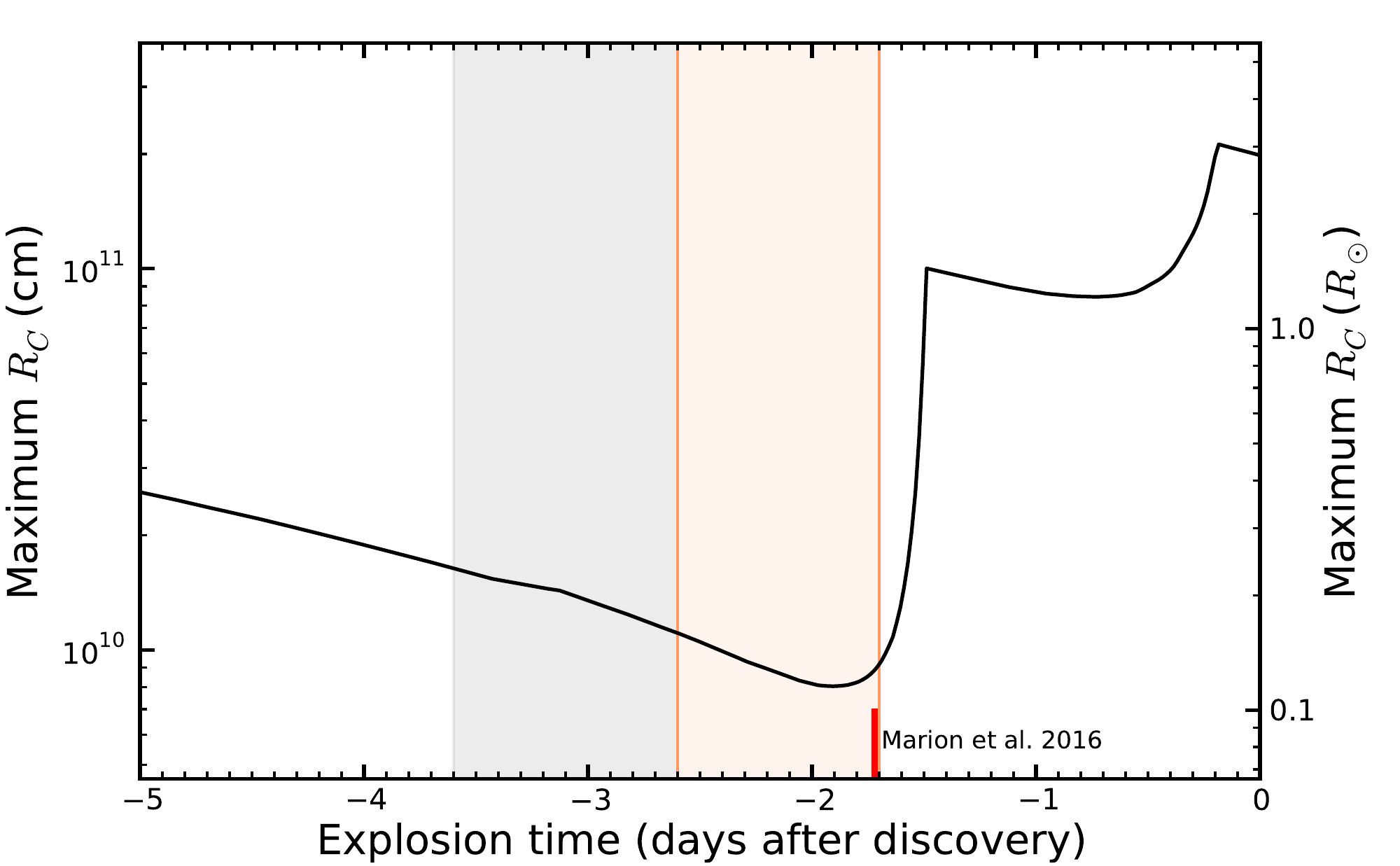}

	\caption{Maximum allowed companion radius as a function of explosion time assuming Roche-lobe overflow and a 1 \msun{} companion.  The orange region shows the conservative estimate for $t_{\textrm{first}}$ we adopt (see Section~\ref{sec:lcfit}). The red tick mark shows the $t_{\textrm{first}}$ fit by \citet{marion16}. The gray region shows a conservative range for which a possible dark phase might have occurred. We then adopt the worst constraint over both these ranges as our conservative constraint on the progenitor system.  These constraints on $R_\textrm{c}$ uses our MASTER upper limit, which was not part of the \citet{marion16} analysis.  
	}
	\label{fig:companion}
\end{figure}

We constrain the radius of a possible non-degenerate companion by using the analytic models provided by \citet{kasen10}. We note that this is conservative because we are ignoring light from the SN itself and attributing all flux to an interaction with a companion.  Because the time of explosion ($t_{\textrm{exp}}$) is uncertain, we explore a variety of $t_{\textrm{exp}}$ and their corresponding constraints on $R_\textrm{c}$. The general procedure is to choose an $t_{\textrm{exp}}$ and then find the maximum $R_\textrm{c}$ that is consistent with the early-time data, assuming the secondary filled its Roche lobe. This introduces a weak dependence on the mass ratio of the binary, and we simply assume the primary and companion masses of 1.4 and $1\,\msun$, respectively. The results are summarized in Figure~\ref{fig:companion}, which shows that $R_\textrm{c} < \rmax{} \rsun{}$ if $t_{\textrm{exp}}$ is within our conservative estimate of $t_{\textrm{first}}$. While $t_{\textrm{exp}}$ and $t_{\textrm{first}}$ have sometimes been used interchangeably in the literature, differences arise because of a possible dark phase between the explosion and when the SN first starts to brighten \citep{hachinger13, piro14, piro16}. \citet{piro16} showed that even in extreme cases dark phases last $< 2$ days, and more realistically last $\lesssim$ 1 day.  Thus, considering one day before our earliest estimate for $t_{\textrm{first}}$ is conservative and a possible companion must still be $R_\textrm{c} < \rmaxdark{}$ \rsun{}.  These estimates assume a viewing angle of 15 degrees, which is more representative than the 0 degree viewing angle shown in the plots of \citet{marion16}.  However, as shown by \citet{kasen10} and noted by \citet{marion16}, the strengthen of the expected signature from a given companion can change by over an order of magnitude based on viewing angle, so a constraint on any given SN Ia needs to be carefully considered.

Although there are a broad range of potential progenitor scenarios, non-degenerate companions can roughly be broken into roughly three main classes: 
1) red giant companions with $R_\textrm{c} > 100 \,\rsun$ that are similar to the observed binaries RS Oph and T CrB (e.g., \citealp{hachisu01}), 
2) helium star companions with $R_\textrm{c} \sim 0.3-6\,\rsun$ that are like the helium-nova system V445 Pup (e.g., \citealp{kato08}),
or 3) main-sequence or subgiant companions with $R_\textrm{c} \sim 0.4-4\,\rsun$ that are similar to U Sco (e.g., \citealp{thoroughgood01}). 
Our limits on $R_\textrm{c}$ rule all three of these for all but the most unfavorable viewing angles.

\section{Conclusions}
\label{sec:sumamry}

Even though SN~2012cg is one of the best-studied \sneia{} to date, the nature of its progenitor is still being debated as we discussed in the introduction.  Specifically, it is difficult to reconcile the claimed early-time light curve detection of a $\sim 6$ \msun{} main-sequence companion from \citet{marion16} with the \citet{maguire16} deep, late-time limit ($< 0.005$  \msun{}) on the amount of low-velocity, solar-metallicity material \citep{liu16}. In this study, we add a new late-time H$\alpha$ constraints, a new X-ray constraint, and a new early-time light curve constraint using a timely prediscovery MASTER observation.  

We obtained a high-signal-to-noise spectrum of SN~2012cg 286 days after $t_{B {\rm max}}$ using LBT/MODS.  We then used the models of \citet{mattila05} to place an upper limit of \MassLimit{} on the amount of solar-abundance material in SN~2012cg. This limit agrees with the limit of $< 0.005$  \msun{} reported by \citet{maguire16}, although we view our limit as more conservative. Determining the expected late-time H$\alpha$ emission in \sneia{} spectra requires difficult radiative transfer calculations, and theoretical questions remain. Most important, the excitation of H$\alpha$ emission by gamma-ray deposition should be modeled as a function of time while varying the amount and velocity profile of stripped material.  These issues notwithstanding, \citet{maguire16} and our mass limit pose a significant challenge to SD models, firmly ruling out MS and RG companions.

We analyzed \swift{} X-ray data for SN~2012cg and place limits on the mass-loss rate and wind velocity from its progenitor system.
Because of X-ray emission from the host galaxy, our \swift{}/XRT limits are not very constraining, with $L<1.4\times 10^{39}\,\rm{erg\,s^{-1}}$ (0.3-10 keV).  For a wind-like CSM, this implies that $\dot M<10^{-6}\,\rm{M_{\sun}yr^{-1}}$ for $v_\textrm{w}=100\,\rm{km\,s^{-1}}$.  For this limit, only a fraction of symbiotic progenitor systems are excluded, and we would not expect to detect X-ray signatures from MS or subgiant companions. 

Finally, we carefully reanalyze a prediscovery MASTER observation in combination with published photometry \citep{silverman12,marion16} to estimate $t_{\textrm{first}}$.  We then looked for the signature expected from an interaction between the SN ejecta and a possible SD companion.  Assuming that $t_{\textrm{exp}}$ is within our range for $t_{\textrm{first}}$ of \toconservdw{} to \toconservup{} days from discovery, we constrain the radius of a companion to be $R_\textrm{c} < \rmax{}$ \rsun{}.  Additionally, \citet{piro16} recently showed that a dark phase between $t_{\textrm{exp}}$ and $t_{\textrm{first}}$ would last less than 1 day is all but the most extreme models.  If $t_{\textrm{exp}}$ is with a day of our conservative estimate of $t_{\textrm{first}}$, then our limit weakens but still constrains  $R_\textrm{c} < \rmaxdark{}$ \rsun{}.  These estimates assume a viewing angle of 15 degrees and rule out a red giant, a helium star, a main sequence, or a subgiant companion for all but the most unfavorable viewing angles.

Our limits on $R_\textrm{c}$ are in agreement with those presented in \citet{silverman12}.  However, they are in stark contrast with the \citet{marion16} claim of a blue early-time excess in the optical light curve consistent with the SN ejecta interacting with a $\sim 6$ \msun{} main-sequence companion.  While there are variations between our analyses, the main differences leading to this discrepancy are:  
\\* 1) \citet{marion16} did not include the MASTER prediscovery observation from \citet{lipunov12ATel}.  This observation proves to be extremely constraining.  
\\* 2) In this study, we fit unfiltered observations as a separate ``filter''.   \citet{marion16} instead shifted the unfiltered ROTSE light curve to match a $t^2$ model fit to $B$-band observations between 10 to 14 days before $t_{B {\rm max}}$.  However, if the ROTSE light curve rises less steeply than $B$-band light curve, then using the ROTSE observations shifted to match $B$ at later times would lead to a spurious excess in flux at early times.  These "overluminous" ROTSE observations would then appear to support a $\sim 6$ \msun{} main-sequence companion, exactly as seen in Figure 4 and 6 of \citet{marion16}.
\\* 3) \citet{marion16} do not include the $r$, $R$, $i$, and $I$ light curves when fitting $t_{\textrm{first}}$ and searching for a signature of a companion.  We use these longer wavelength observations where there is no signature from a potential companion seen. Additionally, once removing the ROTSE points as argued in (2), the $V$- and $B$-band plots in upper two panels of Figure 6 in \citet{marion16} also both show no evidence of a large companion. 
\\* 4) In Figures 4 and 6 of \citet{marion16}, the first $u$-band \swift{} observation appears to show excess flux.  However, in Section \ref{sec:marionlcs} we caution the reader to be careful when interpreting the $u$-band observations for two reasons.  First, SNe Ia have strong Ca II H\&K lines (3945 \AA) that rapidly change which affect this this $u$-band. Second, the Sloan $u$ and \swift{} $u$ are not the same band pass and a careful $S$-correction is required to attempt to translate one into the other. This correction was not preformed in \citet{marion16} and is beyond the scope of this study.
\\* 5) In this study, we do not use the \swift{} UV observations because there are not enough early-time observations to robustly fit the rise.  Figure~5 of \citet{marion16} shows that there is excess flux in the first two \swift{} UV epochs when compared to a $t^{3.6}$ power law.  However, this $3.6$ power-law index was derived from the $u$-band light curve, not the UV light curve.  \citet{shappee16a} found that $uvw2$ and $uvw1$ follow a significantly shallower power law than $u$ band for ASASSN-14lp.  While SN~2012cg and ASASSN-14lp exhibit difference rise properties, if the UV is also intrinsically shallower for SN~2012cg, then fitting \swift{} UV observations with a steeper power law at later times would naturally cause a spurious excess flux at earlier times.  
\\* 6) \citet{marion16} find that SN~2012cg rises slowly between the first two \swift{} UV epochs. However, the slope between the first and second \swift{} epochs appears to be inconsistent with the ground-based photometry at optical wavelengths, and this trend increases going to bluer wavelengths (see Figure~\ref{fig:earlyfit}).  This might point to a problem with the first or second epochs of \swift{} observations.  We reran the Swift photometry ourselves but found no obvious reason why the UV photometry should be wrong.  However, in this study we have shown that the longer-wavelength light curves rule out the \citet{kasen10} models for the shock interaction between the SN ejecta and a companion.  Thus we are left with three possibilities: 
i) There is significant diversity in the UV rise of SNe Ia intrinsic to their explosions. We have already seen extreme variability between SNe Ia in the UV near max (e.g., \citealp{brown10, wang12b, brown15b, foley16} where theoretical studies predict metallicity to have a substantial affect (e.g., \citealp{lentz00, sauer08, walker12}). Thus, it would not be surprising if the same were true for the early-time UV properties of SNe Ia.
ii) Our understanding of the interaction between the SN ejecta and a companion is not complete.
iii) We are seeing the interaction between the SN ejecta and nearby material in a different geometry and/or distribution than a companion.  Perhaps it is the debris from the WD-WD merger \citep{pakmor12, schwab12, shen12, piro16}?
\\* 7) Finally, in this study, instead of assuming a fixed $t_{\textrm{exp}}$, we estimate our uncertainty on $t_{\textrm{first}}$ and allow for the expected range of possible dark times between  $t_{\textrm{first}}$ and $t_{\textrm{exp}}$ when determining $R_\textrm{c}$ limits.  This process allows us to quantify the effects of a possible dark phase. 

The light curves of SNe Ia within days of explosion are sensitive tools that enable us to constrain the physical conditions present in the progenitor systems of SNe Ia immediately prior to explosion.  However, there are multiple physical processes and geometries which potentially imprint themselves on these light curves.  Additionally, careful radiation transfer calculations are needed to understand the detailed effects that a given set of physical conditions would have on the early time light curves.  Thus, we should be careful not to over-interpret limited data, especially limited in wavelength, as evidence for one physical interpretation.  Additionally, while this study, \citet{silverman12}, and \citet{marion16} assume a power law rise for the early-time light curve of SN~2012cg, other studies have found that the early-light curve for some SNe~Ia are more complicated (e.g., \citealp{zheng13, zheng14, goobar15, im15}).  Additional theoretical and observation work is needed to understand the diversity of behaviors that SNe Ia exhibit in their first few days.    

In summary, we do not detect stripped hydrogen-rich material, and we place firm limits on the radius of a possible companion.  When our constraints are combined with other studies of SN~2012cg's progenitor system (e.g., \citealp{chomiuk12ATEL, graur12ATEL, silverman12, maguire13, chomiuk16, graur16, liu16, maguire16}), the available observations strongly favor a DD progenitor for SN~2012cg.


In the last 7 years we have seen an explosion of companion constraints from nearby, bright SN that have been observed extremely early:
SN~2009ig ($< 6$ \msun{}; \citealp{foley12}),
SN~2011fe ($< 0.1 - 0.25$ \rsun{}; \citealp{bloom12, goobar15}),
KSN~2011a ($< 2$ \msun{}; \citealp{olling15}),
KSN~2011b ($< 2$ \msun{}; \citealp{olling15}),
SN~2012cg ($< \rmax{} - \rmaxdark{}$ \rsun{}; this paper),
SN~2013dy ($< 0.35$ \rsun{}; \citealp{zheng13}),
SN~2014J ($\lesssim 0.25 - 4$ \rsun{}; \citep{goobar15, siverd15},
ASASSN-14lp ($< 0.34 − 11$ \rsun{}; \citealp{shappee16a}), 
and SN~2015F ($< 1.0$ \rsun{}; \citealp{im15}).
These discoveries are a testament to large, nearby transient surveys (e.g., ASAS-SN, \citealp{shappee14}; LOSS, \citealp{filippenko01}; PTF, \citealp{law09}).
Similar to \citet{bianco11}, all these observations easily rule out of giant companions for Type Ia Supernovae.  Additionally, there are 3 SNe (SN~2011fe, SN~2012cg, SN~2013dy) where the limits rule out main-sequence or subgiant companions and 2 SNe (SN~2011fe, SN~2012cg) where the limits are strong enough to rule out He star companions.  And yet, the signatures of an interaction with a sizable companion star have not been unambiguously seen for any normal SN Ia.  Assuming the 4 best-constrained SNe are representative of the entire population, the fraction of SNe Ia with $R_\textrm{c} > 1$ \rsun{} is $< 25$\% is $< 25$\% with 1-sigma confidence and $<53$\% with 2-sigma confidence.  We will present a more thorough statistical analysis in a future work (Shappee et al., in prep.).

\acknowledgments

We thank Chris Kochanek, Maria Drout, Jennifer van Saders, Josh Simon, Ryan Foley, Louisa Diodato, Tom Holoien, Laura Chomiuk, and Mark Phillips for discussions and encouragement.  

B.S. is supported by NASA through Hubble Fellowship grant HF-51348.001 awarded by the Space Telescope Science Institute, which is operated by the Association of Universities for Research in Astronomy, Inc., for NASA, under contract NAS 5-26555. KZS is supported by NSF grants AST-1515876 and AST-1515927. R.M. acknowledges partial support from the James Arthur Fellowship at NYU and the Research Corporation for Science Advancement during the completion of this project.

This paper used data obtained with the MODS spectrographs built with funding from NSF grant AST-9987045 and the NSF Telescope System Instrumentation Program (TSIP), with additional funds from the Ohio Board of Regents and the Ohio State University Office of Research.  The $LBT$ is an international collaboration among institutions in the United States, Italy and Germany. $LBT$ Corporation partners are: The Ohio State University, and The Research Corporation, on behalf of The University of Notre Dame, University of Minnesota and University of Virginia; The University of Arizona on behalf of the Arizona university system; Istituto Nazionale di Astrofisica, Italy; $LBT$ Beteiligungsgesellschaft, Germany, representing the Max-Planck Society, the Astrophysical Institute Potsdam, and Heidelberg University.  

Funding for the SDSS and SDSS-II has been provided by the Alfred P. Sloan Foundation, the Participating Institutions, the National Science Foundation, the U.S. Department of Energy, the National Aeronautics and Space Administration, the Japanese Monbukagakusho, the Max Planck Society, and the Higher Education Funding Council for England. The SDSS Web Site is http://www.sdss.org/

This research has made use of the NASA/IPAC Extragalactic Database (NED) which is operated by the Jet Propulsion Laboratory, California Institute of Technology, under contract with the National Aeronautics and Space Administration.  This research has made use of NASA's Astrophysics Data System Bibliographic Services.  IRAF is distributed by the National Optical Astronomy Observatory, which is operated by the Association of Universities for Research in Astronomy (AURA) under a cooperative agreement with the National Science Foundation.  This research made use of Astropy, a community-developed core Python package for Astronomy (Astropy Collaboration, 2013).


\bibliographystyle{apj}

\end{document}